\documentstyle[amsfonts,12pt]{article} 
\begin{document} 

\author{Gunn Quznetsov \\ 
quznets@geocities.com} 
\title{LOGIC, TRUTH AND PROBABILITY} 
\date{August 25, 1998} 
\maketitle 

\begin{abstract} 
The propositional logic is generalized on the real numbers field. The 
logical analog of the Bernoulli independent tests scheme is constructed. The 
variant of the nonstandard analysis is adopted for the definition of the 
logical function, which has all properties of the classical probability 
function. The logical analog of the Large Number Law is deduced from 
properties of this function. 
\end{abstract} 

\tableofcontents 

\section{INTRODUCTION} 

There is the evident nigh affinity between the classical probability 
function and the Boolean function of the classical propositional logic. 
These functions are differed by the range of value, only. That is if the 
range of values of the Boolean function shall be expanded from the 
two-elements set $\left\{ 0;1\right\} $ to the segment $\left[ 0;1\right] $ 
of the real numeric axis then the logical analog of the Bernoulli Large 
Number Law \cite{BRN} can be deduced from the logical axioms. And if the 
range of values of such function shall be expanded to the segment of some 
suitable variant of the hyperreal numeric axis then this theorem shall 
insert some new nonstandard purport for the notion ''truth''. These topics 
is considered in this article. 

\section{THE NONSTANDARD NUMBERS} 

Let us consider the set ${\bf N}$ of natural numbers. 

{\bf Definition 2.1:} The $n${\it -part-set} ${\bf S}$ of ${\bf N}$ is 
defined recursively as follows: 

1) ${\bf S}_1=\left\{ 1\right\} $; 

2) ${\bf S}_{\left( n+1\right) }={\bf S}_n\cup \left\{ n+1\right\} $. 

{\bf Definition 2.2: }If ${\bf S}_n$ is the $n$-part-set of ${\bf N}$ and $%
{\bf A}\subseteq {\bf N}$ then $\left\| {\bf A}\cap {\bf S}_n\right\| $ is 
the quantity elements of the set ${\bf A}\cap {\bf S}_n$, and if 

\[ 
\varpi _n\left( {\bf A}\right) =\frac{\left\| {\bf A}\cap {\bf S}_n\right\| }%
n\mbox{,} 
\] 

then $\varpi _n\left( {\bf A}\right) $ is {\it the frequency} of the set $%
{\bf A}$ on the $n$-part-set ${\bf S}_n$. 

{\bf Theorem 2.1:} 

1) $\varpi _n({\bf N})=1$; 

2) $\varpi _n(\emptyset )=0$; 

3) $\varpi _n({\bf A})+\varpi _n({\bf N}-{\bf A})=1$; 

4) $\varpi _n({\bf A}\cap {\bf B})+\varpi _n({\bf A}\cap ({\bf N}-{\bf B}%
))=\varpi _n({\bf A})$. 

{\bf Definition 2.3: }If ''$\lim $'' is the Cauchy-Weierstrass ''limit'' 
then let us denote: 

\[ 
{\bf \Phi ix=}\left\{ {\bf A}\subseteq {\bf N}|\lim_{n\rightarrow \infty 
}\varpi _n({\bf A})=1\right\} \mbox{.} 
\] 

{\bf Theorem 2.2: }${\bf \Phi ix}$ is the filter \cite{DVS}, i.e.: 

1) ${\bf N}\in {\bf \Phi ix}$, 

2) $\emptyset \notin {\bf \Phi ix}$, 

3) if ${\bf A}\in {\bf \Phi ix}$ and ${\bf B}\in {\bf \Phi ix}$ then $({\bf A%
}\cap {\bf B})\in {\bf \Phi ix}$ ; 

4) if ${\bf A}\in {\bf \Phi ix}$ and ${\bf A}\subseteq {\bf B}$ then ${\bf B}%
\in {\bf \Phi ix}$. 

In the following text we shall adopt to our topics the definitions and the 
proofs of the Robinson Nonstandard Analysis \cite{DVS2}: 

{\bf Definition 2.4:} The sequences of the real numbers $\left\langle 
r_n\right\rangle $ and $\left\langle s_n\right\rangle $ are {\it Q-equivalent%
} (denote: $\left\langle r_n\right\rangle \sim \left\langle s_n\right\rangle 
$) if 

\[ 
\left\{ n\in {\bf N}|r_n=s_n\right\} \in {\bf \Phi ix}\mbox{.} 
\] 

{\bf Theorem 2.3:} If ${\bf r}$,${\bf s}$,${\bf u}$ are the sequences of the 
real numbers then 

1) ${\bf r}\sim {\bf r}$, 

2) if ${\bf r}\sim {\bf s}$ then ${\bf s}\sim {\bf r}$; 

3) if ${\bf r}\sim {\bf s}$ and ${\bf s}\sim {\bf u}$ then ${\bf r}\sim {\bf %
u}$. 

{\bf Definition 2.5:} {\it The Q-number} is the set of the Q-equivalent 
sequences of the real numbers, i.e. if $\widetilde{a}$ is the Q-number and $%
{\bf r}\in \widetilde{a}$ and ${\bf s}\in \widetilde{a}$, then ${\bf r}\sim 
{\bf s};$ and if ${\bf r}\in \widetilde{a}$ and ${\bf r}\sim {\bf s}$ then $%
{\bf s}\in \widetilde{a}$. 

{\bf Definition 2.6:} The Q-number $\widetilde{a}$ is {\it the standard 
Q-number} $a$ if $a$ is some real number and the sequence $\left\langle 
r_n\right\rangle $ exists, for which: $\left\langle r_n\right\rangle \in 
\widetilde{a}$ and 

\[ 
\left\{ n\in {\bf N}|r_n=a\right\} \in {\bf \Phi ix}\mbox{.} 
\] 

{\bf Definition 2.7:} The Q-numbers $\widetilde{a}$ and $\widetilde{b}$ are 
{\it the equal Q-numbers} (denote: $\widetilde{a}=\widetilde{b}$) if a $%
\widetilde{a}\subseteq \widetilde{b}$ and $\widetilde{b}\subseteq \widetilde{%
a}$. 

{\bf Theorem 2.4: }Let ${\frak f}(x,y,z)$ be a function, which has got the 
domain in ${\bf R}\times {\bf R}\times {\bf R}$, has got the range of values 
in ${\bf R}$ (${\bf R}$ is the real numbers set). 

Let $\left\langle y_{1,n}\right\rangle $ , $\left\langle 
y_{2,n}\right\rangle $ , $\left\langle y_{3,n}\right\rangle $ , $%
\left\langle z_{1,n}\right\rangle $ , $\left\langle z_{2,n}\right\rangle $ , 
$\left\langle z_{3,n}\right\rangle $ be any sequences of real numbers. 

In this case if $\left\langle z_{i,n}\right\rangle \sim \left\langle 
y_{i,n}\right\rangle $ then $\left\langle {\frak f}(y_{1,n},y_{2,n},y_{3,n})%
\right\rangle \sim \left\langle {\frak f}(z_{1,n},z_{2,n},z_{3,n})\right%
\rangle $. 

{\bf Definition 2.8:} Let us denote: $Q{\bf R}$ is the set of the Q-numbers. 
~ 

{\bf Definition 2.9: }The function $\widetilde{{\frak f}}$, which has got 
the domain in $Q{\bf R}\times Q{\bf R}\times Q{\bf R}$, has got the range of 
values in $Q{\bf R}$, is {\it the Q-extension of the function} ${\frak f}$, 
which has got the domain in ${\bf R}\times {\bf R}\times {\bf R}$, has got 
the range of values in ${\bf R}$, if the following condition is accomplished: 

Let $\left\langle x_n\right\rangle $ ,$\left\langle y_n\right\rangle $ ,$%
\left\langle z_n\right\rangle $ be any sequences of real numbers. In this 
case: if 

$\left\langle x_n\right\rangle \in \widetilde{x}$, $\left\langle 
y_n\right\rangle \in \widetilde{y}$, $\left\langle z_n\right\rangle \in 
\widetilde{z}$, $\widetilde{u}=\widetilde{{\frak f}}\left( \widetilde{x},%
\widetilde{y},\widetilde{z}\right) $, 

then 

$\left\langle {\frak f}\left( x_n,y_n,z_n\right) \right\rangle \in 
\widetilde{u}$. 

{\bf Theorem 2.5:} For all functions ${\frak f}$, which have the domain in $%
{\bf R}\times {\bf R}\times {\bf R}$, have the range of values in ${\bf R}$, 
and for all real numbers $a$, $b$, $c$, $d$: if $\widetilde{{\frak f}}$ is 
the Q-extension of ${\frak f}$; $\widetilde{a}$, $\widetilde{b}$, $%
\widetilde{c}$, $\widetilde{d}$ are standard Q-numbers $a$, $b$, $c$, $d$, 
then: 

if $d={\frak f}(a,b,c)$ then $\widetilde{d}=\widetilde{{\frak f}}(\widetilde{%
a},\widetilde{b},\widetilde{c})$ and vice versa. ~ 

By this Theorem: if $\widetilde{{\frak f}}$ is the Q-extension of the 
function ${\frak f}$ then the expression ''$\widetilde{{\frak f}}(\widetilde{%
x},\widetilde{y},\widetilde{z})$'' will be denoted as ''${\frak f}(%
\widetilde{x},\widetilde{y},\widetilde{z})$'' and if $\widetilde{u}$ is the 
standard Q-number then the expression ''$\widetilde{u}$'' will be denoted as 
''$u$''. 

{\bf Theorem 2.6:} If for all real numbers $a$, $b$, $c$: 

\[ 
\varphi (a,b,c)=\psi (a,b,c) 
\] 

then for all Q-numbers $\widetilde{x}$, $\widetilde{y}$, $\widetilde{z}$: 

\[ 
\varphi (\widetilde{x},\widetilde{y},\widetilde{z})=\psi (\widetilde{x},%
\widetilde{y},\widetilde{z})\mbox{.} 
\] 

{\bf Theorem 2.7:} If for all real numbers $a$, $b$, $c$: 

\[ 
{\frak f}\left( a,\varphi (b,c)\right) =\psi (a,b,c) 
\] 

then for all Q-numbers $\widetilde{x}$, $\widetilde{y}$, $\widetilde{z}$: 

\[ 
{\frak f}\left( \widetilde{x},\varphi (\widetilde{y},\widetilde{z})\right) 
=\psi (\widetilde{x},\widetilde{y},\widetilde{z})\mbox{.} 
\] 

{\bf Consequences from Theorems 2.6 and 2.7:} \cite{DVS3}: For all Q-numbers 
$\widetilde{x}$, $\widetilde{y}$, $\widetilde{z}$: 

${\bf \Phi }${\bf 1:} $(\widetilde{x}+\widetilde{y})=(\widetilde{y}+%
\widetilde{x})$, 

${\bf \Phi }${\bf 2:} $(\widetilde{x}+(\widetilde{y}+\widetilde{z}))=((%
\widetilde{x}+\widetilde{y})+\widetilde{z})$, 

${\bf \Phi }${\bf 3:} $(\widetilde{x}+0)=\widetilde{x}$, 

${\bf \Phi }${\bf 5:} $(\widetilde{x}\cdot \widetilde{y})=(\widetilde{y}%
\cdot \widetilde{x})$, 

${\bf \Phi }${\bf 6:} $(\widetilde{x}\cdot (\widetilde{y}\cdot \widetilde{z}%
))=((\widetilde{x}\cdot \widetilde{y})\cdot \widetilde{z})$, 

${\bf \Phi 7}${\bf : }$(\widetilde{x}\cdot 1)=\widetilde{x}$, 

${\bf \Phi }${\bf 10:} $(\widetilde{x}\cdot (\widetilde{y}+\widetilde{z}))=((%
\widetilde{x}\cdot \widetilde{y})+(\widetilde{x}\cdot \widetilde{z}))$. 

{\bf Theorem 2.8: }${\bf \Phi }${\bf 4:} For every Q-number $\widetilde{x}$ 
the Q-number $\widetilde{y}$ exists, for which: 

$(\widetilde{x}+\widetilde{y})=0$. 

{\bf Theorem 2.9: }${\bf \Phi 9}${\bf :} There is not that $0=1$. 

{\bf Definition 2.10:} The Q-number $\widetilde{x}$ is {\it Q-less} than the 
Q-number $\widetilde{y}$ (denote: $\widetilde{x}<\widetilde{y}$) if the 
sequences $\left\langle x_n\right\rangle $ and $\left\langle 
y_n\right\rangle $ of real numbers exist, for which: $\left\langle 
x_n\right\rangle \in \widetilde{x}$, $\left\langle y_n\right\rangle \in 
\widetilde{y}$ and 

\[ 
\left\{ n\in {\bf N}|x_n<y_n\right\} \in {\bf \Phi ix}\mbox{.} 
\] 

{\bf Theorem 2.10:} For all Q-numbers $\widetilde{x}$, $\widetilde{y}$, $%
\widetilde{z}$: \cite{DVS4} 

${\bf \Omega 1}$: there is not that $\widetilde{x}<\widetilde{x}$; 

${\bf \Omega 2}$: if $\widetilde{x}<\widetilde{y}$ and $\widetilde{y}<%
\widetilde{z}$ then $\widetilde{x}<\widetilde{z}$; 

${\bf \Omega 4}$: if $\widetilde{x}<\widetilde{y}$ then $(\widetilde{x}+%
\widetilde{z})<(\widetilde{y}+\widetilde{z})$; 

${\bf \Omega 5}$: if $0<\widetilde{z}$ and $\widetilde{x}<\widetilde{y}$, 
then $(\widetilde{x}\cdot \widetilde{z})<(\widetilde{y}\cdot \widetilde{z})$; 

${\bf \Omega 3}^{\prime }$: if $\widetilde{x}<\widetilde{y}$ then there is 
not, that $\widetilde{y}<\widetilde{x}$ or $\widetilde{x}=\widetilde{y}$ and 
vice versa; 

${\bf \Omega 3}^{\prime \prime }$: for all standard Q-numbers $x$, $y$, $z$: 
$x<y$ or $y<x$ or $x=y$. 

{\bf Theorem 2.11: }${\bf \Phi }${\bf 8:} If $0<|\widetilde{x}|$ then the 
Q-number $\widetilde{y}$ exists, for which $(\widetilde{x}\cdot \widetilde{y}%
)=1$. 

Thus, Q-numbers are fulfilled to all properties of real numbers, except $%
\Omega $3 \cite{DVS5}. The property $\Omega $3 is accomplished by some weak 
meaning ($\Omega $3' and $\Omega $3''). 

{\bf Definition 2.11:} The Q-number $\widetilde{x}$ is {\it the 
infinitesimal Q-number} if the sequence of real numbers $\left\langle 
x_n\right\rangle $ exists, for which: $\left\langle x_n\right\rangle \in 
\widetilde{x}$ and for all positive real numbers $\varepsilon $: 

\[ 
\left\{ n\in {\bf N}||x_n|<\varepsilon \right\} \in {\bf \Phi ix}\mbox{.} 
\] 

Let the set of all infinitesimal Q-numbers be denoted as $I$. 

{\bf Definition 2.12:} The Q-numbers $\widetilde{x}$ and $\widetilde{y}$ are 
t{\it he infinite closed Q-numbers} (denote: $\widetilde{x}\approx 
\widetilde{y}$) if $|\widetilde{x}-\widetilde{y}|=0$ or $|\widetilde{x}-%
\widetilde{y}|$ is infinitesimal. 

{\bf Definition 2.13}: The Q-number $\widetilde{x}$ is {\it the infinite 
Q-number} if the sequence $\left\langle r_n\right\rangle $ of real numbers 
exists, for which $\left\langle r_n\right\rangle \in \widetilde{x}$ and for 
every natural number $m$: 

\[ 
\left\{ n\in {\bf N}|m<r_n\right\} \in {\bf \Phi ix}\mbox{.} 
\] 

\section{THE CLASSICAL LOGIC.} 

{\bf Definition 3.1: }The sentence $\ll \Theta \gg $ is {\it the true 
sentence} if and only if $\Theta $ \footnote{%
Perhaps, the definition of the truth sentence belongs to A.Tarsky.}. 

For example: the sentence $\ll $it rains$\gg $ is the true sentence if and 
only if it rains. 

{\bf Definition 3.2: }The sentence $\ll \Theta \gg $ is {\it the false 
sentence} if and only if it is not that $\Theta $. 

{\bf Definition 3.3: }The sentences $A$ and $B$ are {\it equal} ($A=B$) if $%
A $ is true if and only if $B$ is true. 

Hereinafter we use the usual notions of the classical propositional logic 
\cite{M}. 

{\bf Definition 3.4: }The sentence $C$ is {\it the conjunction} of the 
sentences $A$ and $B$ ($C=\left( A\wedge B\right) $) , if $C$ is true if and 
only if $A$ is true and $B$ is true. 

{\bf Definition 3.5: }The sentence $C$ is {\it the negation} of the sentence 
$A$ ( $C=\overline{A}$), if $C$ is true if and only if $A$ is false. 

{\bf Theorem 3.1: } 

1) $(A\wedge A)=A$; 

2) $(A\wedge B)=(B\wedge A)$; 

3) $(A\wedge (B\wedge C))=((A\wedge B)\wedge C)$; 

4) if $T$ is the true sentence then for every sentence $A$: $(A\wedge T)=A$. 

{\bf Definition 3.6:} Each function ${\frak g}$, which has got the domain in 
the set of the sentences, has got the range of values on the two-elements 
set $\left\{ 0;1\right\} $, is {\it the Boolean function} if: 

1) for every sentence $A$: ${\frak g}\left( \overline{A}\right) =1-$ ${\frak %
g}\left( A\right) $; 

2) for all sentences $A$ and $B$: ${\frak g}\left( A\wedge B\right) ={\frak g%
}\left( A\right) \cdot {\frak g}\left( B\right) $. 

{\bf Definition 3.7:} The set $\Im $ of the sentences is {\it the basic set} 
if for every element $A$ of this set the Boolean functions ${\frak g}_1$ and 
${\frak g}_2$ exist, for which the following conditions fulfill: 

1) ${\frak g}_1\left( A\right) \neq {\frak g}_2\left( A\right) $; 

2) for each element $B$ of $\Im $: if $B\neq A$ then ${\frak g}_1\left( 
B\right) ={\frak g}_2\left( B\right) $. 

{\bf Definition 3.8:} The set $\left[ \Im \right] $ of the sentences is {\it %
the propositional closure} of the set $\Im $ if the following conditions 
fulfill: 

1) if $A\in \Im $ then $\overline{A}\in \Im $; 

2) if $A\in \Im $ and $B\in \Im $ then $\left( A\wedge B\right) \in \Im $. 

In the following text the elements of $\left[ \Im \right] $ are denoted as 
{\it the }${\it \Im }${\it -sentences}. 

{\bf Definition 3.9:} The $\Im $-sentence $A$ is {\it the tautology} if for 
all Boolean functions ${\frak g}$: 

\[ 
{\frak g}(A)=1\mbox{.} 
\] 

{\bf Definition 3.10:} {\it The disjunction} and {\it the implication} are 
defined by the usual way: 

\[ 
\begin{array}{c} 
\left( A\vee B\right) =\overline{\left( \overline{A}\wedge \overline{B}%
\right) }\mbox{,} \\ 
\left( A\Rightarrow B\right) =\overline{\left( A\wedge \overline{B}\right) }%
\mbox{.} 
\end{array} 
\] 

By this definition and the Definitions 3.4 and 3.5: 

$\left( A\vee B\right) $ is the false sentence if and if, only, $A$ is the 
false sentence and $B$ is the false sentence. 

$\left( A\Rightarrow B\right) $ is the false sentence if and if, only, $A$ 
is the true sentence and $B$ is the false sentence. 

{\bf Definition 3.11: }A $\Im $-sentence is {\it a propositional axiom }\cite 
{M2} if this sequence has got one some amongst the following forms: 

{\bf A1}. $\left( A\Rightarrow \left( B\Rightarrow A\right) \right) $; 

{\bf A2. }$\left( \left( A\Rightarrow \left( B\Rightarrow C\right) \right) 
\Rightarrow \left( \left( A\Rightarrow B\right) \Rightarrow \left( 
A\Rightarrow C\right) \right) \right) $; 

{\bf A3}. $\left( \left( \overline{B}\Rightarrow \overline{A}\right) 
\Rightarrow \left( \left( \overline{B}\Rightarrow A\right) \Rightarrow 
B\right) \right) $. 

{\bf Definition 3.12: } 

The sentence $B$ is obtained from the sentences $\left( A\Rightarrow 
B\right) $ and $A$ by the logic rule ''{\it modus ponens}''. 

{\bf Definition 3.13: }\cite{M21} The array $A_1,A_2,\ldots ,A_n$ of the 
sentences is {\it the propositional deduction} of the sentence $A$ from the 
hypothesis list $\Gamma $ (denote: $\Gamma \vdash A$), if $A_n=A$ and for 
all numbers $l$ ($1\leq l\leq n$): $A_l$ is either the propositional axiom 
or $A_l$ is obtained from some sentences $A_{l-k}$ and $A_{l-s}$ by the 
modus ponens or $A_l\in \Gamma $. 

{\bf Definition 3.14:} The sentence is t{\it he propositional proved sentence%
} if this sentence is the propositional axiom or this sentence is obtained 
from the propositional proved sentences by the modus ponens. 

Hence, if $A$ is the propositional proved sentence then the propositional 
deduction 

\[ 
\vdash A 
\] 

exists. 

{\bf Theorem 3.2: }\cite{M3} If the sentence $A$ is the propositional proved 
sentence then for all Boolean function ${\frak g}$: ${\frak g}\left( 
A\right) =1$. 

{\bf Theorem 3.3: (The completeness Theorem). }\cite{M4} All tautologies are 
the propositional proved sentences.{\bf \ } 

\section{B-FUNCTIONS} 

{\bf Definition 4.1: }Each function ${\frak b}\left( x\right) $, which has 
got the domain in the sentences set, has got the range of values on the 
numeric axis segment $\left[ 0;1\right] $, is named as {\it the B-function} 
if for every sentences $A$ and $B$ the following condition fulfills: 

\[ 
{\frak b}\left( A\wedge B\right) +{\frak b}\left( A\wedge \overline{B}%
\right) ={\frak b}\left( A\right) \mbox{.} 
\] 

{\bf Theorem 4.1:} For each B-function ${\frak b}$: 

1) for every sentences $A$ and $B$: ${\frak b}\left( A\wedge B\right) \leq 
{\frak b}\left( A\right) $; 

2) for every sentence $A$: if $T$ is the true sentence, then ${\frak b}%
\left( A\right) +{\frak b}\overline{A}={\frak b}\left( T\right) $ 

3) for every sentence $A$: if $T$ is the true sentence, then ${\frak b}%
\left( A\right) \leq {\frak b}\left( T\right) $; 

Therefore, if the sentence $C$ exists, for which: ${\frak b}\left( C\right) 
=1,$ and $T$ is the true sentence, then 

\begin{equation} 
{\frak b}\left( T\right) =1\mbox{.}  \label{b2} 
\end{equation} 

Hence, in this case for every sentence $A$: 

\begin{equation} 
{\frak b}\left( A\right) +{\frak b}\left( \overline{A}\right) =1\mbox{.} 
\label{b3} 
\end{equation} 

{\bf Theorem 4.2:} If the sentence $D$ is the propositional proved sentence 
then for all B-functions ${\frak b}$: ${\frak b}\left( D\right) =1$. 

{\bf Theorem 4.3:} 

1) If for all Boolean functions ${\frak g}$: 

\[ 
{\frak g}\left( A\right) =1 
\] 

then for all B-functions ${\frak b}$: 

\[ 
{\frak b}\left( A\right) =1\mbox{.} 
\] 

2) If for all Boolean functions ${\frak g}$: 

\[ 
{\frak g}\left( A\right) =0 
\] 

then for all B-functions ${\frak b}$: 

\[ 
{\frak b}\left( A\right) =0\mbox{.} 
\] 

{\bf Theorem 4.4:} All Boolean functions are the B-functions. 

{\bf Hence, the B-function is the generalization of the logic Boolean 
function. Therefore, the B-function is the logic function, too.} 

{\bf Theorem 4.5:} 

\[ 
{\frak b}\left( A\vee B\right) ={\frak b}\left( A\right) +{\frak b}\left( 
B\right) -{\frak b}\left( A\wedge B\right) \mbox{.} 
\] 

{\bf Definition 4.2:} The sentences $A$ and $B$ are {\it the inconsistent 
sentences for the B-function }${\frak b}$ if 

\[ 
{\frak b}\left( A\wedge B\right) =0\mbox{.} 
\] 

{\bf Theorem 4.6:} If the sentences $A$ and $B$ are the inconsistent 
sentences for the B-function ${\frak b}$ then 

\[ 
{\frak b}\left( A\vee B\right) ={\frak b}\left( A\right) +{\frak b}\left( 
B\right) \mbox{.} 
\] 

{\bf Theorem 4.7:} If ${\frak b}\left( A\wedge B\right) ={\frak b}\left( 
A\right) \cdot {\frak b}\left( B\right) $ then ${\frak b}\left( A\wedge 
\overline{B}\right) ={\frak b}\left( A\right) \cdot {\frak b}\left( 
\overline{B}\right) $. 

{\bf Theorem 4.8:} ${\frak b}\left( A\wedge \overline{A}\wedge B\right) =0$. 

\section{THE INDEPENDENT TESTS} 

{\bf Definition 5.1:} Let ${\frak st}(n)$ be a function, which has got the 
domain on the set of natural numbers and has got the range of values in the 
set of the $\Im $-sentences. 

In this case, the $\Im $-sentence $A$ is {\it the [st]-series of the range }$%
r${\it \ with the V- number }$k$ if $A$, $r$ and $k$ fulfill to some one 
amongst the following conditions: 

1) $r=1$ and $k=1$, $A={\frak st}\left( 1\right) $ or $k=0$, $A=\overline{%
{\frak st}\left( 1\right) }$; 

2) $B$ is the [st]-series of the range $r-1$ with the V-number $k-1$ and 

\[ 
A=\left( B\wedge {\frak st}\left( r\right) \right) \mbox{,} 
\] 

or $B$ is the [st]-series of the range $r-1$ with the V-number $k$ and 

\[ 
A=\left( B\wedge \overline{{\frak st}\left( r\right) }\right) \mbox{.} 
\] 

Let us denote the set of the [st]-series of the range $r$ with the V-number $%
k$ as $[{\frak st}](r,k)$. 

For example, if ${\frak st}\left( n\right) $ is the sentence $B_n$ then the 
sentences: 

$\left( B_1\wedge B_2\wedge \overline{B_3}\right) $, $\left( B_1\wedge 
\overline{B_2}\wedge B_3\right) $, $\left( \overline{B_1}\wedge B_2\wedge 
B_3\right) $ 

are the elements of $[{\frak st}](3,2)$, and $\left( B_1\wedge B_2\wedge 
\overline{B_3}\wedge B_4\wedge \overline{B_5}\right) \in [{\frak st}](5,3)$. 

{\bf Definition 5.2: }The function ${\frak st}(n)$ is {\it independent} for 
the B-function ${\frak b}$ if for $A$: if $A\in $ $[{\frak st}](r,r)$ then: 

\[ 
{\frak b}\left( A\right) =\prod\limits_{n=1}^r{\frak b}\left( {\frak st}%
\left( n\right) \right) \mbox{.} 
\] 

{\bf Definition 5.3:.}Let ${\frak st}(n)$ be a function, which has got the 
domain on the set of natural numbers and has got the range of values in the 
set of the $\Im $-sentences. 

In this case the sentence $A$ is {\it the [st]-disjunction of the range }$r$%
{\it \ with the V-number }$k$ (denote: ${\frak t}[{\frak st}](r,k)$) if $A$ 
is the disjunction of all elements of $[{\frak st}](r,k)$. 

For example, if ${\frak st}\left( n\right) $ is the sentence $C_n$ then: 

$\left( \overline{C_1}\wedge \overline{C_2}\wedge \overline{C_3}\right) =%
{\frak t}[{\frak st}]\left( 3,0\right) $, 

${\frak t}[{\frak st}]\left( 3,1\right) =\left( \left( C_1\wedge \overline{%
C_2}\wedge \overline{C_3}\right) \vee \left( \overline{C_1}\wedge C_2\wedge 
\overline{C_3}\right) \vee \left( \overline{C_1}\wedge \overline{C_2}\wedge 
C_3\right) \right) $, 

${\frak t}[{\frak st}]\left( 3,2\right) =\left( \left( C_1\wedge C_2\wedge 
\overline{C_3}\right) \vee \left( \overline{C_1}\wedge C_2\wedge C_3\right) 
\vee \left( C_1\wedge \overline{C_2}\wedge C_3\right) \right) $, 

$\left( C_1\wedge C_2\wedge C_3\right) ={\frak t}[{\frak st}]\left( 
3,3\right) $. 

{\bf Definition 5.4:} 

$\nu _r\left[ st\right] \left( A\right) $ is {\it the frequency of the 
sentence }$A$\ {\it \ in the [st]-series of }$r${\it \ independent for the 
B-function }${\frak b}${\it \ tests} if 

1) ${\frak st}(n)$ is independent for the B-function ${\frak b}$, 

2) for all $n$: ${\frak b}\left( {\frak st}\left( n\right) \right) ={\frak b}%
\left( A\right) $, 

3) ${\frak t}[{\frak st}](r,k)=$''$\nu _r\left[ st\right] \left( A\right) 
=\frac kr$''. 

{\bf Theorem 5.1: (the J.Bernoulli formula }\cite{BRN1}{\bf )} If ${\frak st}%
(n)$ is independent for the B-function ${\frak b}$, the real number $p$ 
exists, for which: for all $n$: ${\frak b}\left( {\frak st}\left( n\right) 
\right) =p$, then 

\[ 
{\frak b}\left( {\frak t}\left[ {\frak st}\right] \left( r,k\right) \right) =%
\frac{r!}{k!\cdot \left( r-k\right) !}\cdot p^k\cdot \left( 1-p\right) ^{r-k}%
\mbox{.} 
\] 

{\bf Definition 5.5: }Let ${\frak st}(n)$ be a function, which has got the 
domain on the set of the natural numbers and has got the range of values in 
the set of the $\Im $-sentences. 

In this case the function ${\frak T}[{\frak st}](r,k,l)$, which has got the 
domain in the set of threes of the natural numbers and has got the range of 
values in the set of the $\Im $-sentences, is defined recursively as follows: 

1) ${\frak T}[{\frak st}](r,k,k)={\frak t}[{\frak st}](r,k)$, 

2) ${\frak T}[{\frak st}](r,k,l+1)=({\frak T}[{\frak st}](r,k,l)\vee {\frak t%
}[{\frak st}](r,l+1))$. 

{\bf Definition 5.6:} If $a$ and $b$ are a real numbers and $k-1<a\leq k$ 
and $l\leq b<l+1$ then ${\frak T}[{\frak st}](r,a,b)={\frak T}[{\frak st}%
](r,k,l)$. 

{\bf Theorem 5.2:} If $\nu _r\left[ st\right] \left( A\right) $ is the 
frequency of the sentence $A$\ in the [st]-series of $r${\it \ }independent 
for the B-function ${\frak b}$ tests then 

\[ 
{\frak T}[{\frak st}](r,a,b)="\frac ar\leq \nu _r\left[ st\right] \left( 
A\right) \leq \frac br"\mbox{.} 
\] 

{\bf Theorem 5.3:} If ${\frak st}(n)$ is independent for the B-function $%
{\frak b}$, the real number $p$ exists, for which: for all $n$: ${\frak b}%
\left( {\frak st}\left( n\right) \right) =p$, then 

\[ 
{\frak b}\left( {\frak T}[{\frak st}](r,a,b)\right) =\sum_{a\leq k\leq b}%
\frac{r!}{k!\cdot \left( r-k\right) !}\cdot p^k\cdot \left( 1-p\right) ^{r-k}%
\mbox{.} 
\] 

{\bf Theorem 5.4:} If ${\frak st}(n)$ is independent for the B-function $%
{\frak b}$, the real number $p$ exists, for which: for all $n$: ${\frak b}%
\left( {\frak st}\left( n\right) \right) =p$, then for every positive real 
number $\varepsilon $: 

\[ 
{\frak b}\left( {\frak T}[{\frak st}](r,r\cdot \left( p-\varepsilon \right) 
,r\cdot \left( p+\varepsilon \right) )\right) \geq 1-\frac{p\cdot \left( 
1-p\right) }{r\cdot \varepsilon ^2}\mbox{.} 
\] 

\section{THE PROBABILITY FUNCTION} 

{\bf Definition 6.1:} The sequences of the sentences $\left\langle 
A_n\right\rangle $ and $\left\langle B_n\right\rangle $ are {\it Q-equivalent%
} (denote: $\left\langle A_n\right\rangle \sim \left\langle B_n\right\rangle 
$) if 

\[ 
\left\{ n\in {\bf N}|A_n=B_n\right\} \in {\bf \Phi ix}\mbox{.} 
\] 

{\bf Definition 6.2:} {\it The Q-sentence} is the set of the Q-equivalent 
sequences of the sentences, i.e. if $\widetilde{A}$ is the Q-number and $%
{\bf B}\in \widetilde{A}$ and ${\bf C}\in \widetilde{A}$, then ${\bf B}\sim 
{\bf C};$ and if ${\bf B}\in \widetilde{A}$ and ${\bf B}\sim {\bf C}$ then $%
{\bf C}\in \widetilde{A}$. 

{\bf Definition 6.3:} The Q-sentence $\widetilde{A}$ is {\it the standard 
Q-sentence} $A$ if $A$ is some sentence and sequence $\left\langle 
B_n\right\rangle $ exists, for which: $\left\langle B_n\right\rangle \in 
\widetilde{A}$ and 

\[ 
\left\{ n\in {\bf N}|B_n=A\right\} \in {\bf \Phi ix}\mbox{.} 
\] 

{\bf Definition 6.4:} The Q-sentences $\widetilde{A}$ and $\widetilde{B}$ 
are {\it the equal Q-sentences} (denote: $\widetilde{A}=\widetilde{B}$) if a 
$\widetilde{A}\subseteq \widetilde{B}$ and $\widetilde{B}\subseteq 
\widetilde{A}$. 

{\bf Definition 6.5:} The function $\widetilde{{\frak f}}$,which has got the 
domain in the set of the Q-sentences, has got the range of values in the set 
of the Q-numbers, is {\it the Q-extension of the function} ${\frak f}$, 
which has got the domain in the set of the sentences, has the range of 
values in the set of the real numbers, if the following condition is 
accomplished: 

if $\left\langle B_n\right\rangle \in \widetilde{B}$ and $\widetilde{{\frak f%
}}(\widetilde{B})=\widetilde{x}$ then $\left\langle {\frak f}%
(B_n)\right\rangle \in \widetilde{x}$. ~ 

{\bf Definition 6.6:} The function $\widetilde{{\frak T}}$, which has got 
the domain in $Q{\bf R}\times Q{\bf R}\times Q{\bf R}$, has got the range of 
values in the set of the Q-sentences, is {\it the Q-extension of the 
function }${\frak T}$, which has got the domain in ${\bf R}\times {\bf R}%
\times {\bf R}$, has the range of values in the set of the sentences, if the 
following condition is fulfilled: 

if $\left\langle x_n\right\rangle \in \widetilde{x}$, $\left\langle 
y_n\right\rangle \in \widetilde{y}$, $\left\langle z_n\right\rangle \in 
\widetilde{z}$, $\widetilde{u}=\widetilde{{\frak T}}\left( \widetilde{x},%
\widetilde{y},\widetilde{z}\right) $, then $\left\langle {\frak T}\left( 
x_n,y_n,z_n\right) \right\rangle \in \widetilde{u}$. 

{\bf Theorem 6.1:} Let $x$, $y$, $z$ be the standard Q-numbers, $B$ be the 
standard Q-sentence, $\widetilde{{\frak f}}$ be the Q-extension of the 
function ${\frak f}$, which has got the domain in the set of the sentences 
and has got the range of values in the set of the real numbers. ~ 

Let $\widetilde{{\frak T}}$ be the Q-extension of the function ${\frak T}$, 
which has got the domain in ${\bf R}\times {\bf R}\times {\bf R}$ and has 
got the range of values in the set of the sentences. ~~ 

In this case: 

1) if $\widetilde{{\frak f}}(B)=x$ then ${\frak f}(B)=x$ and vice versa; 

2) if $\widetilde{{\frak T}}(x,y,z)=\widetilde{B}$ then ${\frak T}(x,y,z)=B$ 
and vice versa; 

By this Theorem: if $\widetilde{{\frak f}}$ and $\widetilde{{\frak T}}$ are 
the Q-extensions of the functions ${\frak f}$ and ${\frak T}$ 
correspondingly, then the expressions of the type: ''$\widetilde{{\frak f}}$%
'' and ''$\widetilde{{\frak T}}$'' shall be denote as ''${\frak f}$'' and ''$%
{\frak T}$'' correspondingly. And if $\widetilde{B}$ is the standard 
Q-sentence $B$ then the expressions of the type: ''$\widetilde{B}$'' will be 
denote as ''$B$''. 

{\bf Theorem 6.2:} For all Q-numbers $\widetilde{\varepsilon }$ and $%
\widetilde{r}$, for all functions ${\frak st}\left( n\right) $, independent 
for the B-function ${\frak b}$: 

if $\widetilde{\varepsilon }>0$ and a real number $p$ exists, for which: for 
all natural $n$: ${\frak st}\left( n\right) =p$, then 

\[ 
{\frak b}\left( {\frak T}[{\frak st}](\widetilde{r},\widetilde{r}\cdot 
\left( p-\widetilde{\varepsilon }\right) ,\widetilde{r}\cdot \left( p+%
\widetilde{\varepsilon }\right) )\right) \geq 1-\frac{p\cdot \left( 
1-p\right) }{\widetilde{r}\cdot \widetilde{\varepsilon }^2}\mbox{.} 
\] 

{\bf Theorem 6.3:} If $\widetilde{r}$ is the infinite Q-number then for all 
real positive numbers $\varepsilon $, for all functions ${\frak st}\left( 
n\right) $, independent for the B-function ${\frak b}$: 

if a real number $p$ exists, for which: for all natural $n$: ${\frak st}%
\left( n\right) =p$, then 

\[ 
{\frak b}\left( {\frak T}[{\frak st}](\widetilde{r},\widetilde{r}\cdot 
\left( p-\widetilde{\varepsilon }\right) ,\widetilde{r}\cdot \left( p+%
\widetilde{\varepsilon }\right) )\right) \approx 1\mbox{.} 
\] 

{\bf Definition 6.7:} The sentence ''$\Theta $'' is {\it the almost 
authentic sentence} if it is real, that $\Theta $. 

For example: 

Certain raffle is kept one million raffle tickets. The Big prize falls to 
the single ticket of this raffle. All tickets are sold. You have got one 
ticket. 

In this case, the sentence ''You shall not win the Big prize'' is the almost 
authentic sentence because it is real that you shall not win the Big prize. 
But the sentence ''Someone will win the Big prize.'' is the true sentence by 
the Definition 3.1. 

Hence, all true sentences are the almost authentic sentences, but not all 
almost authentic sentences are the true sentences.{\bf \ } 

{\bf Definition 6.8:} A function ${\frak P}$ is a P-function if ${\frak P}$ 
is a Q-extension of a B-function and the following condition is fulfilled: 

for all Q-sentences $\widetilde{A}$: if ${\frak P}\left( \widetilde{A}%
\right) \approx 1$ then $\widetilde{A}$ is an almost authentic sentence. 

{\bf Theorem 6.4:} If $\widetilde{r}$ is the infinite Q-number, $\nu _{%
\widetilde{r}}\left[ st\right] \left( A\right) $ is the frequency of the 
sentence $A$\ in the [st]-series of $\widetilde{r}${\it \ i}ndependent for 
any P-function ${\frak P}$ tests, then it is real, that for each real 
positive number $\varepsilon $: 

\[ 
\left| \nu _{\widetilde{r}}\left[ st\right] \left( A\right) -{\frak P}\left( 
A\right) \right| <\varepsilon \mbox{.} 
\] 

{\bf Theorem 6.5: }If $\widetilde{r}$ is the infinite Q-number, $\nu _{%
\widetilde{r}}\left[ st\right] \left( A\right) $ is the frequency of the 
sentence $A$\ in the [st]-series of $\widetilde{r}${\it \ i}ndependent for a 
P-function ${\frak P}$ tests, then it is real, that 

\[ 
\nu _{\widetilde{r}}\left[ st\right] \left( A\right) \approx {\frak P}\left( 
A\right) \mbox{.} 
\] 

Therefore, {\bf the function, defined by the Definition 6.8 has got the 
statistical meaning.} That is why I'm name such function as {\it the 
probability function}. 

\section{RESUME} 

The probability function is the extension of the logic B-function. 
Therefore, {\bf the probability is some generalization of the classic 
propositional logic.} 

\section{APPENDIX I. Consistency} 

Let us define the propositional calculus like to (\cite{M}), but the 
propositional forms shall be marked by the script greek letters. 

{\bf Definition C1: }A set $\Re $ of the propositional forms is{\it \ a 
U-world} if: 

1) if $\alpha _1,\alpha _2,\ldots ,\alpha _n\in \Re $ and $\alpha _1,\alpha 
_2,\ldots ,\alpha _n\vdash \beta $ then $\beta \in \Re $, 

2) for all propositional forms $\alpha $: it is not that $(\alpha \&\left( 
\neg \alpha \right) )\in \Re $, 

3) for every propositional form $\alpha $: $\alpha \in \Re $ or $(\neg 
\alpha )\in \Re $. 

{\bf Definition C2: }The sequences of the propositional forms $\left\langle 
\alpha _n\right\rangle $ and $\left\langle \beta _n\right\rangle $ are {\it %
Q-equivalent} (denote: $\left\langle \alpha _n\right\rangle \sim 
\left\langle \beta _n\right\rangle $) if 

\[ 
\left\{ n\in {\bf N}|\alpha _n\equiv \beta _n\right\} \in {\bf \Phi ix}%
\mbox{.} 
\] 

Let us define the notions of {\it the Q-forms} and {\it the Q-extension of 
the functions} for the propositional forms like as in the Definitions 2.5, 
6.2, 2.9, 6.5, 6.6. 

{\bf Definition C3:} The Q-form $\widetilde{\alpha }$ is {\it Q-real} in the 
U-world $\Re $ if the sequence $\left\langle \alpha _n\right\rangle $ of the 
propositional forms exists, for which: $\left\langle \alpha _n\right\rangle 
\in \widetilde{\alpha }$ and 

\[ 
\left\{ n\in {\bf N}|\alpha _n\in \Re \right\} \in {\bf \Phi ix}\mbox{.} 
\] 

{\bf Definition C4: }The set $\widetilde{\Re }$ of the Q-forms is t{\it he 
Q-extension of the U-world }$\Re $ if $\widetilde{\Re }$ is the set of 
Q-forms $\widetilde{\alpha }$, which are Q-real in $\Re $. 

{\bf Definition C5:} The sequence $\left\langle \widetilde{\Re }%
_k\right\rangle $ of the Q-extensions is {\it the S-world}. 

{\bf Definition C6: }The Q-form $\widetilde{\alpha }$ is {\it S-real in the 
S-world }$\left\langle \widetilde{\Re }_k\right\rangle $ if 

\[ 
\left\{ k\in {\bf N}|\widetilde{\alpha }\in \widetilde{\Re }_k\right\} \in 
{\bf \Phi ix}\mbox{.} 
\] 

{\bf Definition C7:} The set ${\bf A}$ (${\bf A}\subseteq {\bf N}$) is {\it %
the regular set} if for every real positive number $\varepsilon $ the 
natural number $n_0$ exists, for which: for all natural numbers $n$ and $m$, 
which are more or equal to $n_0$: 

\[ 
|w_n({\bf A})-w_m({\bf A})|<\varepsilon \mbox{.} 
\] 

{\bf Theorem C1:} If ${\bf A}$ is the regular set and for all real positive $%
\varepsilon $: 

\[ 
\left\{ k\in {\bf N}|w_k({\bf A})<\varepsilon \right\} \in {\bf \Phi ix}%
\mbox{.} 
\] 

then 

\[ 
\lim_{k\rightarrow \infty }w_k({\bf A})=0\mbox{.} 
\] 

{\bf Proof of theTheorem C1:} Let be 

\[ 
\lim_{k\rightarrow \infty }w_k({\bf A})\neq 0\mbox{.} 
\] 

That is the real number $\varepsilon _0$ exists, for which: for every 
natural number $n^{\prime }$ the natural number $n$ exists, for which: 

\[ 
n>n^{\prime }\mbox{ and }w_n({\bf A})>\varepsilon _0. 
\] 

Let $\delta _0$ be some positive real number, for which: $\varepsilon 
_0-\delta _0>0$. Because ${\bf A}$ is the regular set then for $\delta _0$ 
the natural number $n_0$ exists, for which: for all natural numbers $n$ and $%
m$, which are more or equal to $n_0$: 

\[ 
|w_m({\bf A})-w_n({\bf A})|<\delta _0\mbox{.} 
\] 

That is 

\[ 
w_m({\bf A})>w_n({\bf A})-\delta _0\mbox{.} 
\] 

Since $w_n({\bf A})\geq \varepsilon _0$ then $w_m({\bf A})\geq \varepsilon 
_0-\delta _0$. 

Hence, the natural number $n_0$ exists, for which: for all natural numbers $%
m $: if $m\geq n_0$ then $w_m({\bf A})\geq \varepsilon _0-\delta _0$. 

Therefore, 

\[ 
\left\{ m\in {\bf N}|w_m({\bf A})\geq \varepsilon _0-\delta _0\right\} \in 
{\bf \Phi ix}\mbox{.} 
\] 

and by this Theorem condition: 

\[ 
\left\{ k\in {\bf N}|w_k({\bf A})<\varepsilon _0-\delta _0\right\} \in {\bf %
\Phi ix}\mbox{.} 
\] 

Hence, 

\[ 
\left\{ k\in {\bf N}|\varepsilon _0-\delta _0<\varepsilon _0-\delta 
_0\right\} \in {\bf \Phi ix}\mbox{.} 
\] 

That is $\emptyset \notin {\bf \Phi ix}$. It is the contradiction for the 
Theorem 2.2. 

{\bf Definition C8:} Let $\left\langle \widetilde{\Re }_k\right\rangle $ be 
a S-world. 

In this case the function ${\frak W}(\widetilde{\beta })$, which has got the 
domain in the set of the Q-forms, has got the range of values in $Q{\bf R}$, 
is defined as the following: 

If ${\frak W}(\widetilde{\beta })=\widetilde{p}$ then the sequence $%
\left\langle p_n\right\rangle $ of the real numbers exists, for which: $%
\left\langle p_n\right\rangle \in \widetilde{p}$ and 

\[ 
p_n=w_n\left( \left\{ k\in {\bf N}|\widetilde{\beta }\in \widetilde{\Re }%
_k\right\} \right) \mbox{.} 
\] 

{\bf Theorem C2:} If $\left\{ k\in {\bf N}|\widetilde{\beta }\in \widetilde{%
\Re }_k\right\} $ is the regular set and ${\frak W}(\widetilde{\beta }%
)\approx 1$ then $\widetilde{\beta }$ is S-resl in $\left\langle \widetilde{%
\Re }_k\right\rangle $. 

{\bf Proof of the Theorem C2: }Since ${\frak W}(\widetilde{\beta })\approx 1$ 
then by Definitions.2.12 and 2.11: for all positive real $\varepsilon $: 

\[ 
\left\{ n\in {\bf N}|w_n\left( \left\{ k\in {\bf N}|\widetilde{\beta }\in 
\widetilde{\Re }_k\right\} \right) >1-\varepsilon \right\} \in {\bf \Phi ix}%
\mbox{.} 
\] 

Hence, by the point 3 of the Theorem 2.1: for all positive real $\varepsilon 
$: 

\[ 
\left\{ n\in {\bf N}|\left( {\bf N}-w_n\left( \left\{ k\in {\bf N}|%
\widetilde{\beta }\in \widetilde{\Re }_k\right\} \right) \right) 
<\varepsilon \right\} \in {\bf \Phi ix}\mbox{.} 
\] 

Therefore, by the Theorem C1: 

\[ 
\lim_{n\rightarrow \infty }\left( {\bf N}-w_n\left( \left\{ k\in {\bf N}|%
\widetilde{\beta }\in \widetilde{\Re }_k\right\} \right) \right) =0\mbox{.} 
\] 

That is: 

\[ 
\lim_{n\rightarrow \infty }w_n\left( \left\{ k\in {\bf N}|\widetilde{\beta }%
\in \widetilde{\Re }_k\right\} \right) =1\mbox{.} 
\] 

Hence, by Definition.2.3: 

\[ 
\left\{ k\in {\bf N}|\widetilde{\beta }\in \widetilde{\Re }_k\right\} \in 
{\bf \Phi ix}\mbox{.} 
\] 

And by Definition C6: $\widetilde{\beta }$ is S-real in $\left\langle 
\widetilde{\Re }_k\right\rangle $. 

{\bf Theorem C3: }The P-function exists. 

{\bf Proof of the Theorem C3:} By the Theorems C2 and 2.1: ${\frak W}(%
\widetilde{\beta })$ is the P-function in $\left\langle \widetilde{\Re }%
_k\right\rangle $. 

\section{APPENDIX II. Proofs} 

This Appendix contains proofs of the Theorems: 

{\bf Proof of the Theorem 2.1:} This is obvious. 

{\bf Proof of the Theorem 2.2:} From the point 3 of Theorem 2.1: 

\[ 
\lim_{n\rightarrow \infty }\varpi _n({\bf N}-{\bf B})=0\mbox{.} 
\] 

From the point 4 of Theorem 2.1: 

\[ 
\varpi _n({\bf A}\cap ({\bf N}-{\bf B}))\leq \varpi _n({\bf N}-{\bf B})%
\mbox{.} 
\] 

Hence, 

\[ 
\lim_{n\rightarrow \infty }\varpi _n\left( {\bf A}\cap ({\bf N}-{\bf B}%
)\right) =0\mbox{.} 
\] 

Hence, 

\[ 
\lim_{n\rightarrow \infty }\varpi _n\left( {\bf A}\cap {\bf B}\right) 
=\lim_{n\rightarrow \infty }\varpi _n({\bf A})\mbox{.} 
\] 

{\bf Proof of the Theorem 2.3:} By Definition 2.4 from the Theorem 2.2 is 
obvious. 

{\bf Proof of the Theorem 2.4:} Let us denote: 

if $k=1$ or $k=2$ or $k=3$ then 

\[ 
{\bf A}_k=\left\{ n\in {\bf N}|y_{k,n}=z_{k,n}\right\} \mbox{.} 
\] 

In this case by Definition 2.4 for all $k$: 

\[ 
{\bf A}_k\in {\bf \Phi ix}\mbox{.} 
\] 

Because 

\[ 
\left( {\bf A}_1\cap {\bf A}_2\cap {\bf A}_3\right) \subseteq \left\{ n\in 
{\bf N}|{\frak f}(y_{1,n},y_{2,n},y_{3,n})={\frak f}%
(z_{1,n},z_{2,n},z_{3,n})\right\} \mbox{,} 
\] 

then by Theorem 2.2: 

\[ 
\left\{ n\in {\bf N}|{\frak f}(y_{1,n},y_{2,n},y_{3,n})={\frak f}%
(z_{1,n},z_{2,n},z_{3,n})\right\} \in {\bf \Phi ix}\mbox{.} 
\] 

{\bf Proof of the Theorem 2.5:} If $\left\langle r_n\right\rangle \in 
\widetilde{a}$, $\left\langle s_n\right\rangle \in \widetilde{b}$, $%
\left\langle u_n\right\rangle \in \widetilde{c}$, $\left\langle {\frak t}%
_n\right\rangle \in \widetilde{d}$ then by Definition 2.6: 

\[ 
\begin{array}{c} 
\left\{ n\in {\bf N}|r_n=a\right\} \in {\bf \Phi ix}\mbox{,} \\ 
\left\{ n\in {\bf N}|s_n=b\right\} \in {\bf \Phi ix}\mbox{,} \\ 
\left\{ n\in {\bf N}|u_n=c\right\} \in {\bf \Phi ix}\mbox{,} \\ 
\left\{ n\in {\bf N}|t_n=d\right\} \in {\bf \Phi ix}\mbox{.} 
\end{array} 
\] 

1) Let $d={\frak f}(a,b,c)$. 

In this case by Theorem 2.2: 

\[ 
\left\{ n\in {\bf N}|t_n={\frak f}(r_n,s_n,u_n)\right\} \in {\bf \Phi ix}%
\mbox{.} 
\] 

Hence, by Definition 2.4: 

\[ 
\left\langle t_n\right\rangle \sim \left\langle {\frak f}(r_n,s_n,u_n)\right%
\rangle \mbox{.} 
\] 

Therefore by Definition 2.5: 

\[ 
\left\langle {\frak f}(r_n,s_n,u_n)\right\rangle \in \widetilde{d}\mbox{.} 
\] 

Hence, by Definition 2.9: 

\[ 
\widetilde{d}=\widetilde{{\frak f}}(\widetilde{a},\widetilde{b},\widetilde{c}%
)\mbox{.} 
\] 

2) Let $\widetilde{d}=\widetilde{{\frak f}}(\widetilde{a},\widetilde{b},%
\widetilde{c})$. 

In this case by Definition 2.9: 

\[ 
\left\langle {\frak f}(r_n,s_n,u_n)\right\rangle \in \widetilde{d}\mbox{.} 
\] 

Hence, by Definition 2.5: 

\[ 
\left\langle t_n\right\rangle \sim \left\langle {\frak f}(r_n,s_n,u_n)\right%
\rangle \mbox{.} 
\] 

Therefore, by Definition 2.4: 

\[ 
\left\{ n\in {\bf N}|t_n={\frak f}(r_n,s_n,u_n)\right\} \in {\bf \Phi ix}%
\mbox{.} 
\] 

Hence, by the Theorem 2.2: 

\[ 
\left\{ n\in {\bf N}|t_n={\frak f}(r_n,s_n,u_n),r_n=a,s_n=b,u_n=c,t_n=d%
\right\} \in {\bf \Phi ix}\mbox{.} 
\] 

Hence, since this set does not empty, then 

\[ 
d={\frak f}(a,b,c)\mbox{.} 
\] 

{\bf Proof of the Theorem 2.6:} If $\left\langle x_n\right\rangle \in 
\widetilde{x}$, $\left\langle y_n\right\rangle \in \widetilde{y}$, $%
\left\langle z_n\right\rangle \in \widetilde{z}$, $\widetilde{u}=\varphi (%
\widetilde{x},\widetilde{y},\widetilde{z})$, then by Definition 2.9: $%
\left\langle \varphi (x_n,y_n,z_n)\right\rangle \in \widetilde{u}$. 

Because $\varphi (x_n,y_n,z_n)=\psi (x_n,y_n,z_n)$ then $\left\langle \psi 
(x_n,y_n,z_n)\right\rangle \in \widetilde{u}$. 

If $\widetilde{v}=\psi (\widetilde{x},\widetilde{y},\widetilde{z})$ then by 
Definition 2.9: $\left\langle \psi (x_n,y_n,z_n)\right\rangle \in \widetilde{%
v}$, too. 

Therefore, for all sequences $\left\langle t_n\right\rangle $ of real 
numbers: if $\left\langle t_n\right\rangle \in \widetilde{u}$ then by 
Definition 2.5: $\left\langle t_n\right\rangle \sim \left\langle \psi 
(x_n,y_n,z_n)\right\rangle $. 

Hence, $\left\langle t_n\right\rangle \in \widetilde{v}$; and if $%
\left\langle t_n\right\rangle \in \widetilde{v}$ then $\left\langle 
t_n\right\rangle \sim \left\langle \varphi (x_n,y_n,z_n)\right\rangle $; 
hence, $\left\langle t_n\right\rangle \in \widetilde{u}$. 

Therefore, $\widetilde{u}=\widetilde{v}$. 

{\bf Proof of the Theorem 2.7:} Let $\left\langle w_n\right\rangle \in 
\widetilde{w}$, ${\frak f}(\widetilde{x},\widetilde{w})=\widetilde{u}$, $%
\left\langle x_n\right\rangle \in \widetilde{x}$, $\left\langle 
y_n\right\rangle \in \widetilde{y}$, $\left\langle z_n\right\rangle \in 
\widetilde{z}$, $\varphi (\widetilde{y},\widetilde{z})=\widetilde{w}$, $\psi 
(\widetilde{x},\widetilde{y},\widetilde{z})=\widetilde{v}$. 

By the condition of this Theorem: ${\frak f}(x_n,\varphi (y_n,z_n))=\psi 
(x_n,y_n,z_n)$. 

By Definition 2.9: $\left\langle \psi (x_n,y_n,z_n)\right\rangle \in 
\widetilde{v}$, $\left\langle \varphi (x_n,y_n)\right\rangle \in \widetilde{w%
}$, $\left\langle {\frak f}(x_n,w_n)\right\rangle \in \widetilde{u}$. 

For all sequences $\left\langle t_n\right\rangle $ of real numbers: 

1) If $\left\langle t_n\right\rangle \in \widetilde{v}$ then by Definition 
2.5: $\left\langle t_n\right\rangle \sim \left\langle \psi 
(x_n,y_n,z_n)\right\rangle $. 

Hence $\left\langle t_n\right\rangle \sim \left\langle {\frak f}(x_n,\varphi 
(y_n,z_n))\right\rangle $. 

Therefore, by Definition 2.4: 

\[ 
\left\{ n\in {\bf N}|t_n={\frak f}(x_n,\varphi \left( y_n,z_n\right) 
)\right\} \in {\bf \Phi ix} 
\] 

and 

\[ 
\left\{ n\in {\bf N}|w_n=\varphi \left( y_n,z_n\right) \right\} \in {\bf %
\Phi ix}\mbox{.} 
\] 

Hence, by Theorem 2.2: 

\[ 
\left\{ n\in {\bf N}|t_n={\frak f}(x_n,w_n)\right\} \in {\bf \Phi ix}\mbox{.} 
\] 

Hence, by Definition 2.4: 

\[ 
\left\langle t_n\right\rangle \sim \left\langle {\frak f}(x_n,w_n)\right%
\rangle \mbox{.} 
\] 

Therefore, by Definition 2.5: $\left\langle t_n\right\rangle \in \widetilde{u%
}$. 

2) If $\left\langle t_n\right\rangle \in \widetilde{u}$ then by Definition 
2.5: $\left\langle t_n\right\rangle \sim \left\langle {\frak f}%
(x_n,w_n)\right\rangle $. 

Because $\left\langle w_n\right\rangle \sim \left\langle \varphi 
(y_n,z_n)\right\rangle $ then by Definition 2.4: 

\[ 
\left\{ n\in {\bf N}|t_n={\frak f}(x_n,w_n)\right\} \in {\bf \Phi ix}\mbox{,} 
\] 

\[ 
\left\{ n\in {\bf N}|w_n=\varphi \left( y_n,z_n\right) \right\} \in {\bf %
\Phi ix}\mbox{.} 
\] 

Therefore, by Theorem 2.2: 

\[ 
\left\{ n\in {\bf N}|t_n={\frak f}(x_n,\varphi \left( y_n,z_n\right) 
)\right\} \in {\bf \Phi ix}\mbox{.} 
\] 

Hence, by Definition 2.4: 

\[ 
\left\langle t_n\right\rangle \sim \left\langle {\frak f}(x_n,\varphi 
(y_n,z_n))\right\rangle \mbox{.} 
\] 

Therefore, 

\[ 
\left\langle t_n\right\rangle \sim \left\langle \psi 
(x_n,y_n,z_n)\right\rangle \mbox{.} 
\] 

Hence, by Definition 2.5: $\left\langle t_n\right\rangle \in \widetilde{v}$. 

From above and from 1) by Definition 2.7: $\widetilde{u}=\widetilde{v}$. 

{\bf Proof of the Theorem 2.8: }If $\left\langle x_n\right\rangle \in 
\widetilde{x}$ then $\widetilde{y}$ is the Q-number, which contains $%
\left\langle -x_n\right\rangle $. 

{\bf Proof of the Theorem 2.9:} is obvious from Definition 2.6 and 
Definition 2.7. 

{\bf Proof of the Theorem 2.10:} is obvious from Definition 2.10 by the 
Theorem 2.2. 

{\bf Proof of the Theorem 2.11:} If $\left\langle x_n\right\rangle \in 
\widetilde{x}$ then by Definition 2.10: if 

\[ 
{\bf A}=\left\{ n\in {\bf N}|0<\left| x_n\right| \right\} 
\] 

then ${\bf A}\in {\bf \Phi ix}$. 

In this case: if for the sequence $\left\langle y_n\right\rangle $ : if $%
n\in {\bf A}$ then $y_n=1/x_n$ 

- then 

\[ 
\left\{ n\in {\bf N}|x_n\cdot y_n=1\right\} \in {\bf \Phi ix}\mbox{.} 
\] 

{\bf Proof of the Theorem 3.1: }It is obviously. 

{\bf Proof of the Theorem 3.2: }in \cite{M3}. 

{\bf Proof of the Theorem 3.3: }in \cite{M4}. 

{\bf Proof of the Theorem 4.1:} 

1) It is obviously. 

2) From the points 4 and 2 of the Theorem 3.1: ${\frak b}\left( T\wedge 
A\right) +{\frak b}\left( T\wedge \overline{A}\right) ={\frak b}\left( 
A\right) +{\frak b}\left( \overline{A}\right) $. 

3) It is obviously. 

{\bf Proof of the Theorem 4.2: } 

If $D$ is A1 then by Definition.3.10: 

\[ 
{\frak b}\left( D\right) ={\frak b}\left( \overline{\left( A\wedge \overline{%
\overline{\left( B\wedge \overline{A}\right) }}\right) }\right) \mbox{.} 
\] 

By (\ref{b3}): 

\[ 
{\frak b}\left( D\right) =1-{\frak b}\left( A\wedge \overline{\overline{%
\left( B\wedge \overline{A}\right) }}\right) \mbox{.} 
\] 

By the Definition 4.1 and the Theorem 3.1: 

\[ 
\begin{array}{c} 
{\frak b}\left( D\right) =1-{\frak b}\left( A\right) +{\frak b}\left( 
A\wedge \overline{\left( B\wedge \overline{A}\right) }\right) \mbox{,} \\ 
{\frak b}\left( D\right) =1-{\frak b}\left( A\right) +{\frak b}\left( 
A\right) -{\frak b}\left( A\wedge \left( B\wedge \overline{A}\right) \right) %
\mbox{,} \\ 
{\frak b}\left( D\right) =1-{\frak b}\left( \left( A\wedge B\right) \wedge 
\overline{A}\right) \mbox{,} \\ 
{\frak b}\left( D\right) =1-{\frak b}\left( A\wedge B\right) +{\frak b}%
\left( \left( A\wedge B\right) \wedge A\right) \mbox{,} \\ 
{\frak b}\left( D\right) =1-{\frak b}\left( A\wedge B\right) +{\frak b}%
\left( \left( A\wedge A\right) \wedge B\right) \mbox{,} \\ 
{\frak b}\left( D\right) =1-{\frak b}\left( A\wedge B\right) +{\frak b}%
\left( A\wedge B\right) \mbox{.} 
\end{array} 
\] 

The proof is similar for the rest propositional axioms . 

Let for all B-function ${\frak b}$: ${\frak b}(A)=1$ and ${\frak b}%
(A\Rightarrow D)=1$. 

By Definition.3.10: 

\[ 
{\frak b}\left( A\Rightarrow D\right) ={\frak b}\left( \overline{A\wedge 
\overline{D}}\right) \mbox{.} 
\] 

By (\ref{b3}): 

\[ 
{\frak b}\left( A\Rightarrow D\right) =1-{\frak b}\left( A\wedge \overline{D}%
\right) \mbox{.} 
\] 

Hence, 

\[ 
{\frak b}\left( A\wedge \overline{D}\right) =0\mbox{.} 
\] 

By Definition.4.1: 

\[ 
{\frak b}\left( A\wedge \overline{D}\right) ={\frak b}\left( A\right) -%
{\frak b}\left( A\wedge D\right) \mbox{.} 
\] 

Hence, 

\[ 
{\frak b}\left( A\wedge D\right) ={\frak b}\left( A\right) =1\mbox{.} 
\] 

By Definition.4.1 and the Theorem 3.1: 

\[ 
{\frak b}\left( A\wedge D\right) ={\frak b}\left( D\right) -{\frak b}\left( 
D\wedge \overline{A}\right) =1\mbox{.} 
\] 

Therefore, for all B-function ${\frak b}$: 

\[ 
{\frak b}\left( D\right) =1\mbox{.} 
\] 

{\bf Proof of the Theorem 4.3: } 

1) This just follows from the preceding Theorem and from the Theorem 3.3. 

2) If for all Boolean functions ${\frak g}$: ${\frak g}\left( A\right) =0$, 
then by the Definition 3.6: ${\frak g}\left( \overline{A}\right) =1$ . 
Hence, by the point 1 of this Theorem: for all B-function ${\frak b}$: $%
{\frak b}\left( \overline{A}\right) =1$. By (\ref{b3}): ${\frak b}\left( 
A\right) =0$. 

{\bf Proof of the Theorem 4.4: }By Definition 3.6: for all Boolean functions 
${\frak g}$: 

${\frak g}\left( A\wedge B\right) +{\frak g}\left( A\wedge \overline{B}%
\right) ={\frak g}\left( A\right) \cdot {\frak g}\left( B\right) +{\frak g}%
\left( A\right) \cdot \left( 1-{\frak g}\left( B\right) \right) ={\frak g}%
\left( A\right) $. 

{\bf Proof of the Theorem 4.5: }By the Definition 3.10 and (\ref{b3}): 

\[ 
{\frak b}\left( A\vee B\right) =1-{\frak b}\left( \overline{A}\wedge 
\overline{B}\right) . 
\] 

By Definition 4.1: 

\[ 
{\frak b}\left( A\vee B\right) =1-{\frak b}\left( \overline{A}\right) +%
{\frak b}\left( \overline{A}\wedge B\right) ={\frak b}\left( A\right) +%
{\frak b}\left( B\right) -{\frak b}\left( A\wedge B\right) \mbox{.} 
\] 

{\bf Proof of the Theorem 4.6: }This just follows from the preceding Theorem 
and Definition. 

{\bf Proof of the Theorem 4.7: }By the Definition 4.1: 

\[ 
{\frak b}\left( A\wedge \overline{B}\right) ={\frak b}\left( A\right) -%
{\frak b}\left( A\wedge B\right) \mbox{.} 
\] 

Hence, 

\[ 
{\frak b}\left( A\wedge \overline{B}\right) ={\frak b}\left( A\right) -%
{\frak b}\left( A\right) \cdot {\frak b}\left( B\right) ={\frak b}\left( 
A\right) \cdot \left( 1-{\frak b}\left( B\right) \right) \mbox{.} 
\] 

Hence, by (\ref{b3}): 

\[ 
{\frak b}\left( A\wedge \overline{B}\right) ={\frak b}\left( A\right) \cdot 
{\frak b}\left( \overline{B}\right) \mbox{.} 
\] 

{\bf Proof of the Theorem 4.8: }By the Definition 4.1 and by the points 2 
and 3 of the Theorem 3.1: 

\[ 
{\frak b}\left( A\wedge \overline{A}\wedge B\right) ={\frak b}\left( A\wedge 
B\right) -{\frak b}\left( A\wedge A\wedge B\right) , 
\] 

hence, by the point 1 of the Theorem 3.1: 

\[ 
{\frak b}\left( A\wedge \overline{A}\wedge B\right) ={\frak b}\left( A\wedge 
B\right) -{\frak b}\left( A\wedge B\right) \mbox{.} 
\] 

{\bf Proof of the Theorem 5.1: }By the Definition 5.2 and the Theorem 4.7: 
if $B\in \left[ {\frak st}\right] \left( r,k\right) $ then: 

\[ 
{\frak b}\left( B\right) =p^k\cdot \left( 1-p\right) ^{r-k}\mbox{.} 
\] 

Since $\left[ {\frak st}\right] \left( r,k\right) $ contains $\frac{r!}{%
k!\cdot \left( r-k\right) !}$ elements then by the Theorems 4.8 and 4.6 this 
Theorem is fulfilled. 

{\bf Proof of the Theorem 5.2: }By the Definition 5.6: the natural numbers $%
r $ and $k$ exist, for which: $k-1<a\leq k$ and $l\leq b<l+1$. 

The induction on $l$: 

1. Let $l=k$. 

In this case by the Definition 5.4: 

\[ 
{\frak T}[{\frak st}](r,k,k)={\frak t}[{\frak st}](r,k)="\nu _r\left[ 
st\right] \left( A\right) =\frac kr"\mbox{.} 
\] 

2. Let $n$ be any natural number. 

The inductive supposition: Let 

\[ 
{\frak T}[{\frak st}](r,k,k+n)="\frac kr\leq \nu _r\left[ st\right] \left( 
A\right) \leq \frac{k+n}r"\mbox{.} 
\] 

By the Definition 5.5: 

\[ 
{\frak T}[{\frak st}](r,k,k+n+1)=({\frak T}[{\frak st}](r,k,k+n)\vee {\frak t%
}[{\frak st}](r,k+n+1))\mbox{.} 
\] 

By the inductive supposition and by the Definition 5.4: 

\[ 
{\frak T}[{\frak st}](r,k,k+n+1)= 
\] 

\[ 
=("\frac kr\leq \nu _r\left[ st\right] \left( A\right) \leq \frac{k+n}r"\vee 
"\nu _r\left[ st\right] \left( A\right) =\frac{k+n+1}r")\mbox{.} 
\] 

Hence, by the Definition 3.10: 

\[ 
{\frak T}[{\frak st}](r,k,k+n+1)="\frac kr\leq \nu _r\left[ st\right] \left( 
A\right) \leq \frac{k+n+1}r"\mbox{.} 
\] 

{\bf Proof of the Theorem 5.3: }This is the consequence from the Theorem 5.1 
by the Theorem 4.6. 

{\bf Proof of the Theorem 5.4:} Because 

\[ 
\sum_{k=0}^r\left( k-r\cdot p\right) ^2\cdot \frac{r!}{k!\cdot \left( 
r-k\right) !}\cdot p^k\cdot \left( 1-p\right) ^{r-k}=r\cdot p\cdot \left( 
1-p\right) 
\] 

then if 

\[ 
J=\left\{ k\in {\bf N}|0\leq k\leq r\cdot \left( p-\varepsilon \right) 
\right\} \cap \left\{ k\in {\bf N}|r\cdot \left( p+\varepsilon \right) \leq 
k\leq r\right\} 
\] 

then 

\[ 
\sum_{k\in J}\frac{r!}{k!\cdot \left( r-k\right) !}\cdot p^k\cdot \left( 
1-p\right) ^{r-k}\leq \frac{p\cdot \left( 1-p\right) }{r\cdot \varepsilon ^2}%
\mbox{.} 
\] 

Hence, by (\ref{b3}) this Theorem is fulfilled. 

{\bf Proof of the Theorem 6.1: } is in common with the Proof of the Theorem 
2.5. 

{\bf Proof of the Theorem 6.2:} This just follows from the preceding Theorem 
and from the Theorem 5.4 and the Definition 2.10. 

{\bf Proof of the Theorem 6.3: }By the Definition.2.13: the sequence $%
\left\langle r_n\right\rangle $ of real numbers exists, for which $%
\left\langle r_n\right\rangle \in \widetilde{r}$ and for every natural 
number $m$: 

\[ 
\left\{ n\in {\bf N}|m<r_n\right\} \in {\bf \Phi ix}\mbox{.} 
\] 

Hence, for all real positive numbers $\delta $: 

\[ 
\left\{ n\in {\bf N}|\frac{p\cdot \left( 1-p\right) }{\widetilde{r}\cdot 
\widetilde{\varepsilon }^2}<\delta \right\} \in {\bf \Phi ix}\mbox{.} 
\] 

Therefore, by the Definitions.2.11: $\frac{p\cdot \left( 1-p\right) }{%
\widetilde{r}\cdot \widetilde{\varepsilon }^2}$ is the infinitesimal. 
Because by the Definition.4.1 ${\frak b}\leq 1$ then by the Definition.2.12: 
from the Theorem 6.2 this Theorem is fulfilled. 

{\bf Proof of the Theorem 6.4:} By the Theorem 6.3: 

\[ 
{\frak P}\left( {\frak T}[{\frak st}](\widetilde{r},\widetilde{r}\cdot 
\left( {\frak P}\left( A\right) -\widetilde{\varepsilon }\right) ,\widetilde{%
r}\cdot \left( {\frak P}\left( A\right) +\widetilde{\varepsilon }\right) 
)\right) \approx 1\mbox{.} 
\] 

By the Definitions 6.8: The sentence 

\[ 
{\frak T}[{\frak st}](\widetilde{r},\widetilde{r}\cdot \left( {\frak P}%
\left( A\right) -\widetilde{\varepsilon }\right) ,\widetilde{r}\cdot \left( 
{\frak P}\left( A\right) +\widetilde{\varepsilon }\right) ) 
\] 

is the almost authentic sentence. 

By the Theorem 5.2: 

\[ 
{\frak T}[{\frak st}](\widetilde{r},\widetilde{r}\cdot \left( {\frak P}%
\left( A\right) -\widetilde{\varepsilon }\right) ,\widetilde{r}\cdot \left( 
{\frak P}\left( A\right) +\widetilde{\varepsilon }\right) )= 
\] 

\[ 
="\left( {\frak P}\left( A\right) -\widetilde{\varepsilon }\right) \leq \nu 
_{\widetilde{r}}\left[ st\right] \left( A\right) \leq \left( {\frak P}\left( 
A\right) +\widetilde{\varepsilon }\right) "\mbox{.} 
\] 

Hence, by the Definition 6.7: It is real, that $\left( {\frak P}\left( 
A\right) -\widetilde{\varepsilon }\right) \leq \nu _{\widetilde{r}}\left[ 
st\right] \left( A\right) \leq \left( {\frak P}\left( A\right) +\widetilde{%
\varepsilon }\right) $. 

{\bf Proof of the Theorem 6.5:} Because from the Theorem 6.4 it is real, 
that for each real positive number $\varepsilon $: 

\[ 
\left| \nu _{\widetilde{r}}\left[ st\right] \left( A\right) -{\frak P}\left( 
A\right) \right| <\varepsilon \mbox{,} 
\] 

then by the Definition 2.11: $\left| \nu _{\widetilde{r}}\left[ st\right] 
\left( A\right) -{\frak P}\left( A\right) \right| $ is the infinitesimal 
Q-number. Hence, by the Definition 2.12 this Theorem is fulfilled.

\end{document}